\documentclass[11pt]{article}
\usepackage{amsmath}
\usepackage[dvips]{graphicx}
\begin{document}        
\title{Exact General Relativistic Rotating Disks Immersed in Rotating Dust
Generated from van Stockum Solutions}
\author{D. Vogt\thanks{e-mail: danielvt@ifi.unicamp.br}\\
Instituto de F\'{\i}sica Gleb Wataghin, Universidade Estadual de Campinas\\
13083-970 Campinas, S.\ P., Brazil
\and
P. S. Letelier\thanks{e-mail: letelier@ime.unicamp.br}\\
Departamento de Matem\'{a}tica Aplicada-IMECC, Universidade Estadual \\
de Campinas 13083-970 Campinas, S.\ P., Brazil}
\maketitle  
\begin{abstract}
A class of exact general relativistic rotating thick disks surrounded by rotating dust is constructed 
by applying the ``displace, cut, fill and reflect'' method on the van Stockum class of metrics. Two 
particular solutions are considered in detail. We find that the disks have annular regions with 
negative energy density and heat flow is present in most of the radial extension of the disk. It is 
shown that thin disks are composed of exotic matter with an equation of state of cosmic strings 
or struts. We also comment the recent relativistic galactic model proposed by Cooperstock and Tieu 
and the presence of an additional disk of exotic matter in their model.  

\textit{Keywords}: Relativistic disks; galaxies; dark matter.   
\end{abstract}
\section{Introduction}
Exact solutions of Einstein's field equations describing stationary or static 
axially symmetric configurations of matter are of great astrophysical interest, since 
the natural shape of an isolated self-gravitating fluid is axially symmetric. In particular, 
disk-like solutions can be used as models of galaxies and accretion disks.
Solutions for static thin disks without radial pressure were first studied by 
Bonnor and Sackfield \cite{Bon68} and Morgan and Morgan \cite{Morg69}, and with 
radial pressure by Morgan and Morgan \cite{Morg70}. Several classes of exact solutions 
of the Einstein field equations corresponding to static thin disks with or without radial 
pressure have been obtained by different authors \cite{Lyn78}--\cite{Gar04}.
Thin rotating disks that can be considered as a source of the Kerr metric were presented 
by Bi\v{c}\'{a}k and Ledvinka \cite{Led93}, while rotating disks with heat flow were studied by 
Gonz\'alez and Letelier \cite{Gonz00}. 
Also thin disks with radial tension \cite{Gonz99}, magnetic fields \cite{Let99} 
and magnetic and electric fields \cite{Katz99} were considered. The nonlinear superposition 
of a disk and a black hole was first obtained by Lemos and Letelier \cite{Lem93}. Perfect
fluid disks with halos \cite{Vogt03} and charged perfect fluid disks \cite{Vogt04} were studied by Vogt and Letelier.
The stability of some General Relativistic thin disk models using a first order perturbation of
the energy-momentum tensor was investigated by Ujevic and Letelier \cite{Uje04}. For a survey on self gravitating relativistic disks, see for instance \cite{Karas04}.

Even though in a first approximation thin disks can be used as useful models, in 
more realistic models the thickness of the disk should be considered. The addition of a 
new dimension may change the dynamical properties of the disk source, e.\ g., its stability. 
Thick static relativistic disks in various coordinate systems were presented by 
Gonz\'alez and Letelier \cite{Gonz04} and Vogt and Letelier \cite{Vogt05a}. General Relativistic versions of some well known Newtonian three-dimensional galactic potential-density pairs were studied by Vogt and Letelier \cite{Vogt05b}. 

In the works cited above an inverse style method was used to solve the Einstein equations, 
i.\ e., the energy-momentum tensor is computed from the metric representing the disk. Another 
approach to generate disks is by solving the Einstein equations given a source (energy-momentum
tensor). Essentialy, they are obtained by solving a Riemann-Hilbert problem and are highly nontrivial. This
has been used by a German group to generate several exact solutions of disks \cite{Klein97}--\cite{Klein03b}. 

Recently Cooperstock and Tieu \cite{Coop05} devised a new approach to modelling galactic dynamics via General 
Relativity. A galaxy is idealised as an axially symmetric uniform rotating fluid without 
pressure (dust). The authors show that the density distribution of the galaxy is 
related to the rotation parameter in a nonlinear way, and then show that several measured rotation curves 
for galaxies are consistent with the mass distribution of matter in flattened disks, therefore 
excluding the need for a massive halo of dark matter. However, their model has been 
criticized by some authors \cite{Korz05,Vogt05c,Garf06}. In particular, 
Korzy\'{n}ski \cite{Korz05} has argued that the model possesses an additional singular rotating thin disk at $z=0$ 
and Vogt and Letelier \cite{Vogt05c} have shown that this singular disk is made of exotic matter either like cosmic strings or
struts with negative energy density. Balasin and Grumiller \cite{Bal06}, using a somewhat different mathematical formulation, 
have shown that is possible to obtain flat rotation curves without unphysical sources at $z=0$, but in turn 
distributional sources may arise on the $z$-axis. Motivated by Cooperstock and Tieu's work, we 
investigate the properties of \emph{thick} rotating disks immersed in rotating dust obtained by using the inverse style method (image method).  

The work is divided as follows. In Sec.\ \ref{sec_stockum} we present the van Stockum \cite{van37} class of 
metrics that will be used to construct the disks and the formalism to calculate their physical variables. 
A class of rotating disks based on the Bonnor \cite{Bon77} solution is presented in Sec.\ \ref{sec_bon}. In 
Sec.\ \ref{sec_quad} another class of rotating disks based on an extension of Bonnor's solution is discussed. 
The thin disk limit and some comments about the Cooperstock and Tieu galactic model are presented in Sec.\ \ref{sec_exotic}. Sec.\ \ref{sec_discuss} is devoted to the discussion of results. 
\section{The van Stockum class of metrics and construction of disks} \label{sec_stockum}

The class of solutions found by van Stockum \cite{van37} describing stationary axially symmetric distributions of dust matter has a line element given by
\begin{equation} \label{eq_line_stockum}
\mathrm{d}s^2= ( \mathrm{d}t-N\mathrm{d}\varphi)^2-r^2\mathrm{d}\varphi^2-
e^{\nu}(\mathrm{d}r^2+\mathrm{d}z^2) \mbox{,}
\end{equation}
where $N$ and $\nu$ are functions of the quasi-cylindrical coordinates $r$ and $z$. For metric Eq.\  
(\ref{eq_line_stockum}), Einstein's field equations $G_{\mu\nu}=(8\pi G/c^4)T_{\mu\nu}$ can be cast as
\begin{subequations}
\begin{align}
& N_{,rr} +N_{,zz}-\frac{N_{,r}}{r}=0 \mbox{,} \label{eq_stockum1} \\
& \nu_{,z} = -\frac{N_{,r}N_{,z}}{r}, \qquad \nu_{,r}=\frac{N_{,z}^2-N_{,r}^2}{2r} \mbox{,} \label{eq_stockum2}\\
& \rho =\frac{c^2}{8\pi Gr^2e^{\nu}}\left( N_{,r}^2+N_{,z}^2 \right) \mbox{,} \label{eq_stockum3}
\end{align}
\end{subequations}
where $\rho$ denotes the energy density of dust. Eq.\ (\ref{eq_stockum1}) can be expressed as 
\begin{equation} \label{eq_Phi}
\Phi_{,rr} +\Phi_{,zz}+\frac{\Phi_{,r}}{r}=0 \mbox{,} 
\end{equation}
by the transformation $N=r\Phi_{,r}$.

In order to obtain a thick disk surrounded by dust, given a solution of Eqs.\ (\ref{eq_stockum1})--(\ref{eq_stockum3}), we apply the 
transformation $z \rightarrow h(z)+b$, where $b$ is a positive constant and $h(z)$ an even function of $z$ such that $h(0)=0$. The parameter $b$ 
controls how much matter is concentrated near the symmetry axis (larger values of $b$ generate smoother distributions of matter); whereas the function 
$h(z)$ dictates the matter distribution along the $z$ direction. This is equivalent to construct a disk by the ``displace, cut, fill and 
reflect'' method that is described at length in \cite{Gonz04}. The nonzero components of the energy-momentum tensor for metric equation (\ref{eq_line_stockum}) are
\begin{subequations}
\begin{align}
T^t_t &=\frac{c^4}{16\pi G e^{\nu}}\left[ \frac{3}{2r^2}(N_{,r}^2+N_{,z}^2)-
(\nu_{,rr}+\nu_{,zz})+\frac{N}{r^2}\left( N_{,rr}+N_{,zz}-\frac{N_{,r}}{r} 
\right) \right] \mbox{,} \label{eq_Ttt} \\
T^t_{\varphi} &=-\frac{c^4}{8\pi G e^{\nu}}\left[ \frac{1}{2}\left( 
N_{,rr}+N_{,zz}-\frac{N_{,r}}{r} \right) \left( 1+\frac{N^2}{r^2}\right)+
\frac{N}{r^2}(N_{,r}^2+N_{,z}^2) \right] \mbox{,}  \\
T^{\varphi}_t &=\frac{c^4}{16\pi G r^2e^{\nu}}\left( N_{,rr}+N_{,zz}-\frac{N_{,r}}{r}\right) \mbox{,}\\
T^{\varphi}_{\varphi} &=-\frac{c^4}{16\pi Ge^{\nu}} \left[ \frac{N}{r^2} \left( 
N_{,rr}+N_{,zz}-\frac{N_{,r}}{r} \right) +\frac{1}{2r^2}( N_{,r}^2+N_{,z}^2)+ 
\nu_{,rr}+\nu_{,zz} \right] \mbox{,} \\
T^r_r &=-T^z_z=\frac{c^4}{32\pi Gr^2e^{\nu}}(N_{,z}^2-N_{,r}^2-2r\nu_{,r}) \mbox{,} \\
T^r_z &=-\frac{c^4}{16\pi Gr^2e^{\nu}}(N_{,r}N_{,z}+r\nu_{,z}) \mbox{.} \label{eq_Trz} 
\end{align}
\end{subequations}
Applying the transformation $z \rightarrow h(z)+b$ and using Eqs.\ (\ref{eq_stockum1})--(\ref{eq_stockum3}) we obtain the following expressions for the components of the energy-momentum tensor of the disk
\begin{subequations}
\begin{align}
T^t_t &=\rho_0c^2+\frac{c^4}{16\pi G e^{\nu}}\left[ h^{\prime\prime} \left( 
\frac{NN_{,h}}{r^2}-\nu_{,h} \right)+(1-h^{\prime 2})\left( \nu_{,hh}-\frac{3N_{,h}^2}{2r^2}
-\frac{NN_{,hh}}{r^2}\right) \right] \mbox{,} \label{eq_Ttt2}\\
T^t_{\varphi} &=-c^2N\rho_0-\frac{c^4}{8\pi G e^{\nu}}\left[ \frac{h^{\prime\prime}N_{,h}}{2}
\left( 1+\frac{N^2}{r^2} \right)-(1-h^{\prime 2}) \times \right. \notag \\
& \left. \left( \frac{NN_{,h}^2}{r^2}+\frac{N_{,hh}}{2}
\left( 1+\frac{N^2}{r^2} \right) \right) \right] \mbox{,} \\
T^{\varphi}_t &=\frac{c^4}{16\pi G r^2e^{\nu}} \left[ h^{\prime\prime}N_{,h}-(1-h^{\prime 2})
N_{,hh} \right] \mbox{,} \\
T^{\varphi}_{\varphi} &=-\frac{c^4}{16\pi G e^{\nu}} \left[ h^{\prime\prime}\left( 
\frac{NN_{,h}}{r^2}+\nu_{,h} \right)-(1-h^{\prime 2})\left( \frac{NN_{,hh}}{r^2}+
\frac{N_{,h}^2}{2r^2}+\nu_{,hh} \right) \right] \mbox{,} \\
T^r_r &=-T^z_z =\frac{c^4}{32\pi Gr^2 e^{\nu}}N_{,h}^2(h^{\prime 2}-1), \qquad T^r_z=0 \mbox{,} \label{eq_Trr_Tzz}
\end{align}
\end{subequations}
where primes indicate differentiation with respect to $z$ and $\rho_0$ is given by Eq.\ 
(\ref{eq_stockum3}) with $ N_{,z}$ replaced by $ N_{,h}$. 

The physical variables of the disk are obtained by solving the eigenvalue problem $T^a_b\xi^b=\lambda\xi^a$, and has the solutions 
\begin{align}
& \lambda_{\pm} =\frac{1}{2}(S\pm \sqrt{D}), \\
S &= T^t_t+ T^{\varphi}_{\varphi}=\rho_0c^2+\frac{c^4}{8\pi G e^{\nu}} \left[ - h^{\prime\prime} 
\nu_{,h}+(1- h^{\prime 2}) \left( \nu_{,hh}-\frac{N_{,h}^2}{2r^2} \right) \right] \mbox{,} \label{eq_S}\\
D &=( T^t_t- T^{\varphi}_{\varphi})^2+4 T^t_{\varphi}T^{\varphi}_t =\left[ \rho_0c^2-
\frac{c^4 N_{,h}^2(1-h^{\prime 2})}{8\pi G r^2e^{\nu}} \right]^2 \notag \\
& -\frac{c^8}{64\pi^2G^2r^2 e^{2\nu}}
\left[h^{\prime\prime}N_{,h}-N_{,hh}(1-h^{\prime 2}) \right]^2 \mbox{,} \label{eq_D}\\
\lambda_r &=T^r_r, \qquad \lambda_z =T^z_z=-T^r_r. 
\end{align}
Defining an orthonormal tetrad $(V^a,W^a,X^a,Y^a)$ with $V^a=N_1(1,\Gamma,0,0)$, 
$W^a=N_2(\Delta,1,0,0)$, $X^a=e^{-\nu/2}(0,0,1,0)$ and $Y^a= e^{-\nu/2}(0,0,0,1)$, where 
$N_1,N_2$ are normalization factors and
\begin{align}
\Gamma &= \begin{cases} 
(\lambda_+-T^t_t)/T^t_{\varphi}, & D \geq 0 \\
(T^t_t-T^{\varphi}_{\varphi})/(2T^t_{\varphi}), & D \leq 0
\end{cases} \\
\Delta &=\begin{cases} 
(\lambda_--T^{\varphi}_{\varphi})/T^{\varphi}_t, & D \geq 0 \\
0, & D \leq 0
\end{cases} \mbox{,}
\end{align}
 the energy-momentum tensor can be written in the canonical form
\begin{equation}
T_{ab}=\rho V_aV_b+P_{\varphi}W_aW_b+\kappa (V_aW_b+W_aV_b)+P_rX_aX_b+P_zY_aY_b 
\mbox{.}
\end{equation}
The energy density $\rho$, the azimuthal stress $P_{\varphi}$, the heat flow function $\kappa$, the 
radial pressure $P_r$ and the tension $P_z$ along $z$ are, respectively, 
\begin{align}
\rho &= \begin{cases} 
\lambda_+/c^2, & D \geq 0 \\
S/(2c^2), & D \leq 0
\end{cases} \label{eq_rho_disc} \\ 
P_{\varphi} &= \begin{cases} 
-\lambda_-, & D \geq 0 \\
-S/2, & D \leq 0
\end{cases} \label{eq_Pphi_disc}\\ 
\kappa &= \begin{cases} 
0, & D \geq 0 \\
\sqrt{D}/2, & D \leq 0
\end{cases} \label{eq_heat_disc} \\ 
P_r &=-T^r_r, \qquad P_z=T^r_r \mbox{.} \label{eq_Pr}
\end{align}

For the functions $h(z)$ we use a class of even polynomials derived in \cite{Vogt05a} 
\begin{equation} \label{eq_h} 
h(z)= \begin{cases}
-z+C, & z \leq -a, \\
Az^2+Bz^{2n+2}, & -a \leq z \leq a, \\
z+C, & z \geq a,
\end{cases}
\end{equation}
with
\[  A =\frac{2n+1-ad}{4na}, \quad B=\frac{ad-1}{4n(n+1)a^{2n+1}}, \quad
C =-\frac{a(2n+1+ad)}{4(n+1)}. \]
Here $n=1,2,\ldots$; $a$ is the half-thickness of the disk and $d$ is the jump of the second derivative on $z=\pm a$. We have 
$|h^{\prime}(z)|=1$ and $h^{\prime \prime}=0$ for $|z|>a$, so Eqs.\ (\ref{eq_Ttt2})--(\ref{eq_Pr}) are reduced to van Stockum's solution outside the disk. For $|z|\leq a$ we always have $|h^{\prime}(z)|\leq 1$ 
and Eqs.\ (\ref{eq_Trr_Tzz}) and (\ref{eq_Pr}) show that we have radial pressures and tensions along $z$. Also when $d=0$ 
$h^{\prime \prime}(|z|=a)=0$ and Eqs.\ (\ref{eq_S})--(\ref{eq_D}) show that the energy density of the 
disk is continuously matched with the surrounding dust cloud on the vertical borders of the disk.
\section{Thick disks from the Bonnor solution} \label{sec_bon}

Bonnor \cite{Bon77} found a simple solution of equations (\ref{eq_stockum1})--(\ref{eq_Phi}) by starting with 
the following solution of Eq.\ (\ref{eq_Phi})
\begin{equation}
\Phi=\frac{2k}{\sqrt{r^2+z^2}},
\end{equation}
where $k$ is a constant. The other functions read
\begin{subequations}
\begin{align}
N &=-\frac{2kr^2}{(r^2+z^2)^{3/2}}, \qquad \nu=\frac{k^2r^2(r^2-8z^2)}{2(r^2+z^2)^4} \mbox{,} \label{eq_bon1}\\
\rho &=\frac{c^2k^2(r^2+4z^2)}{2\pi Ge^{\nu} (r^2+z^2)^4} \mbox{.} \label{eq_bon2}
\end{align}
\end{subequations}
The energy density is singular at the origin. Now applying the transformation $z \rightarrow h(z)+b$ on 
Eqs.\ (\ref{eq_bon1})--(\ref{eq_bon2}) and using Eqs.\ (\ref{eq_S})--(\ref{eq_D}),  
(\ref{eq_Trr_Tzz}) and (\ref{eq_Pr}), we obtain
\begin{subequations}
\begin{align}
D &=\frac{c^8k^4}{4\pi^2G^2e^{2\nu}[r^2+(h+b)^2]^{10}}\left\{ [r^2-2(h+b)^2]^2+9r^2(h+b)^2 
h^{\prime 2}\right\}^2 \notag \\
& -\frac{9c^8k^2r^2}{16\pi^2G^2e^{2\nu}[r^2+(h+b)^2]^7} \left\{ (b+h) 
[r^2+(h+b)^2]h^{\prime \prime} \right. \notag \\
& \left. -(1-h^{\prime 2}) [r^2-4(h+b)^2] \right\}^2 \mbox{,} \label{eq_D_bon} \\
S &= \frac{c^4k^2[r^2+4(h+b)^2]}{2\pi Ge^{\nu} [r^2+(h+b)^2]^4}+\frac{3c^4k^2r^2}{4\pi Ge^{\nu}
 [r^2+(h+b)^2]^6}\left\{ 2 h^{\prime \prime}(h+b)\times \right. \notag \\  
& \left. [r^2-2(h+b)^2][r^2+(h+b)^2]- (1-h^{\prime 2})[2r^4 -27r^2(h+b)^2+31(h+b)^4] \right\} \mbox{,} \label{eq_S_bon} \\
P_r &=-P_z=\frac{9c^4k^2r^2(h+b)^2(1-h^{\prime 2})}{8\pi Ge^{\nu}[r^2+(h+b)^2]^5} \mbox{.} \label{eq_Pr_bon}
\end{align}
\end{subequations}

\begin{figure}
\centering
\includegraphics[scale=0.8]{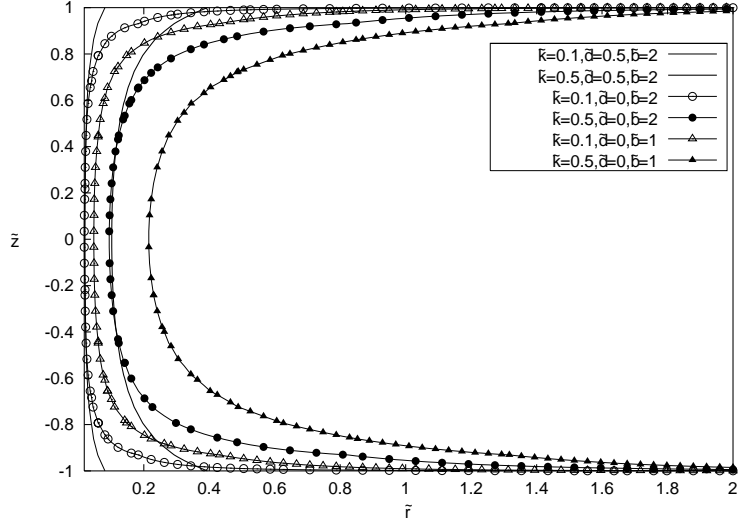}
\caption{Curves of $\tilde{D}=0$ Eq.\ (\ref{eq_D_bon}) as function of $\tilde{r}$ and $\tilde{z}$ 
and some values of $\tilde{k}$, $\tilde{d}$ and $\tilde{b}$. The polynomial $\tilde{h}$ is taken with 
$n=1$.} \label{fig1}
\end{figure}

We nondimensionalise the quantities by defining $\tilde{r}=r/a$, $\tilde{z}=z/a$, $\tilde{h}=h/a$, 
$\tilde{d}=d/a$, $\tilde{k}=k/a^2$, $\tilde{D}=(G^2a^4/c^8)D$, $\tilde{S}=(Ga^2/c^4)S$, 
$\tilde{\rho}=(Ga^2/c^2)\rho$, $\tilde{\kappa}=(Ga^2/c^4)\kappa$ and $\tilde{P}_i=(Ga^2/c^4)P_i$. 
Fig.\ \ref{fig1} displays curves of $\tilde{D}=0$ as function of $\tilde{r}$ and $\tilde{z}$ for the 
polynomial $\tilde{h}$ with $n=1$ and some values of $\tilde{k}$, $\tilde{d}$ and $\tilde{b}$.  On 
the left of each curve the discriminant $\tilde{D}$ is positive.  We observe that increasing the value 
of the parameter $\tilde{k}$ keeping the others fixed, the level curve $\tilde{D}=0$ is moved to the 
right. The same happens when $\tilde{b}$ is decreased and the other parameters are kept fixed.   

\begin{figure}
\centering
\includegraphics[scale=0.5]{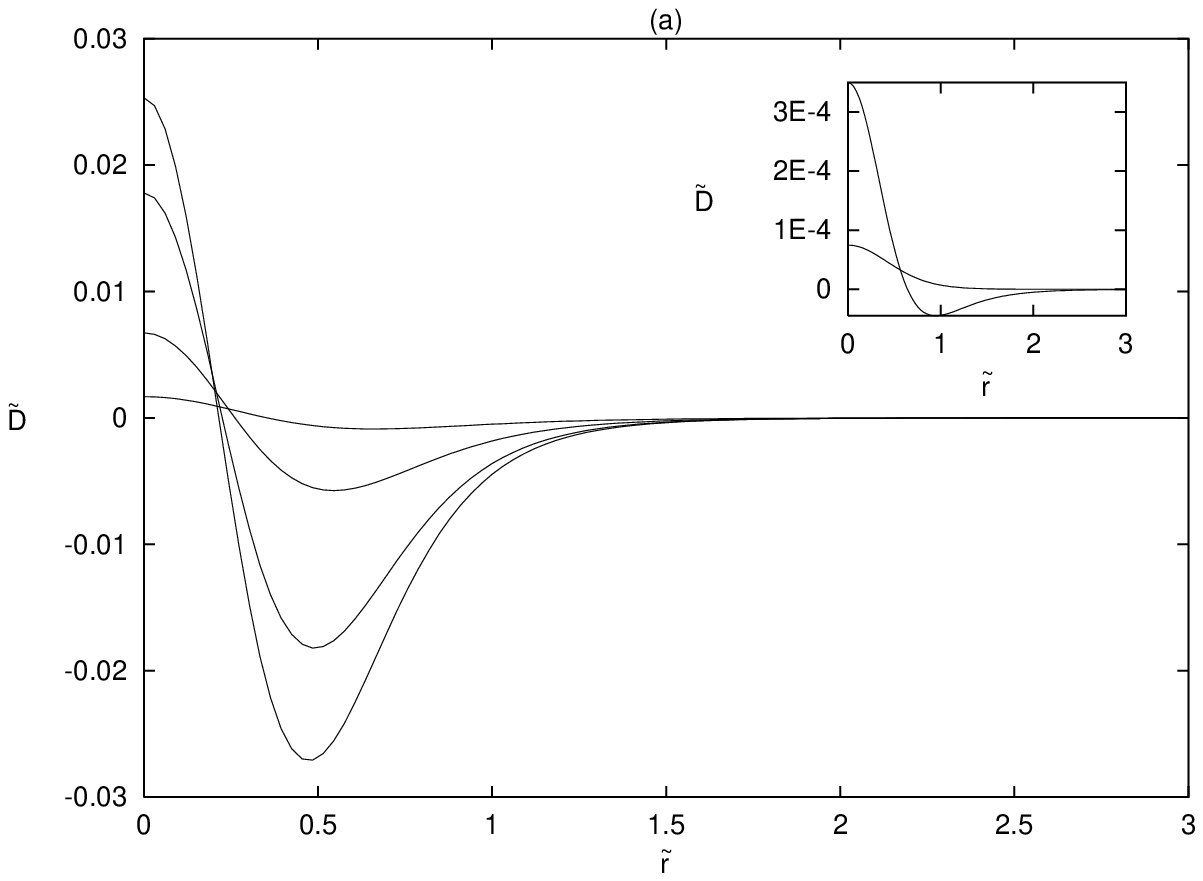}%
\includegraphics[scale=0.5]{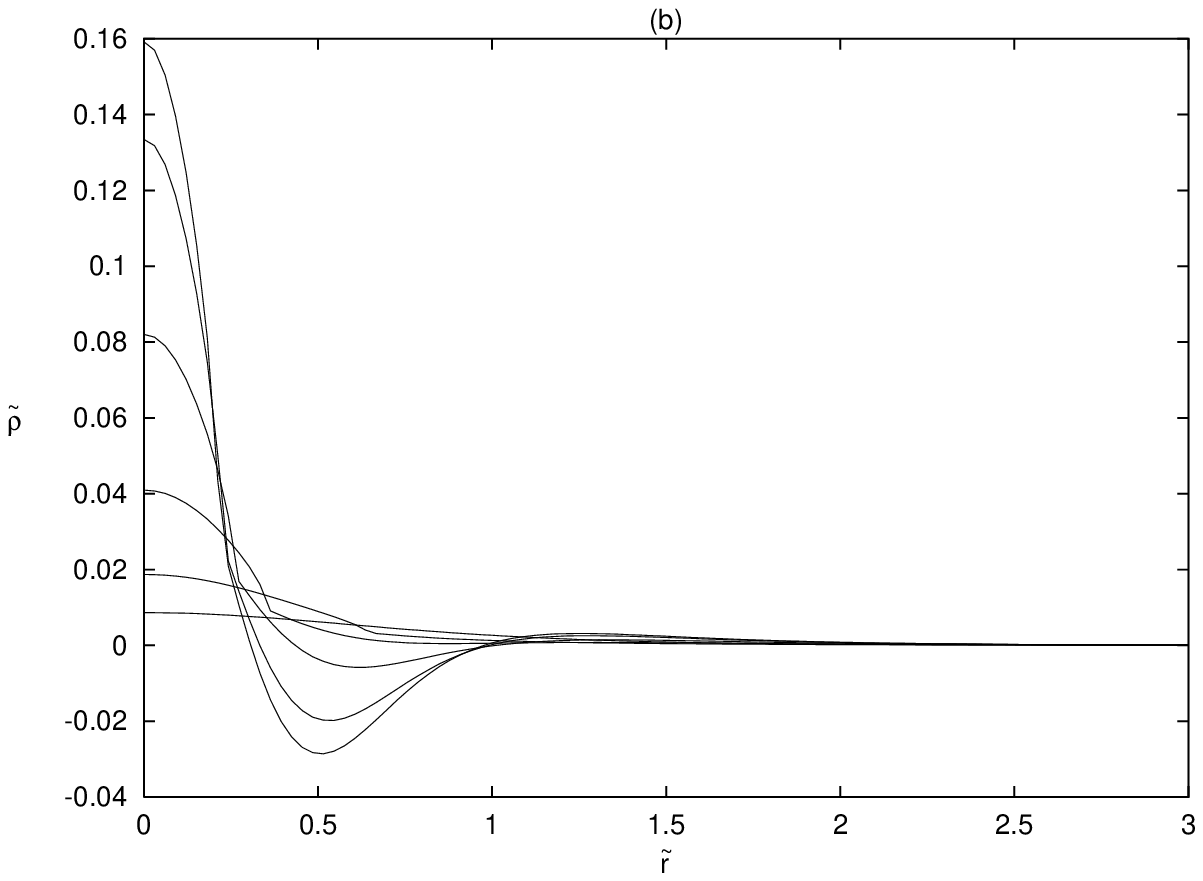}\\
\vspace{0.2cm}
\includegraphics[scale=0.5]{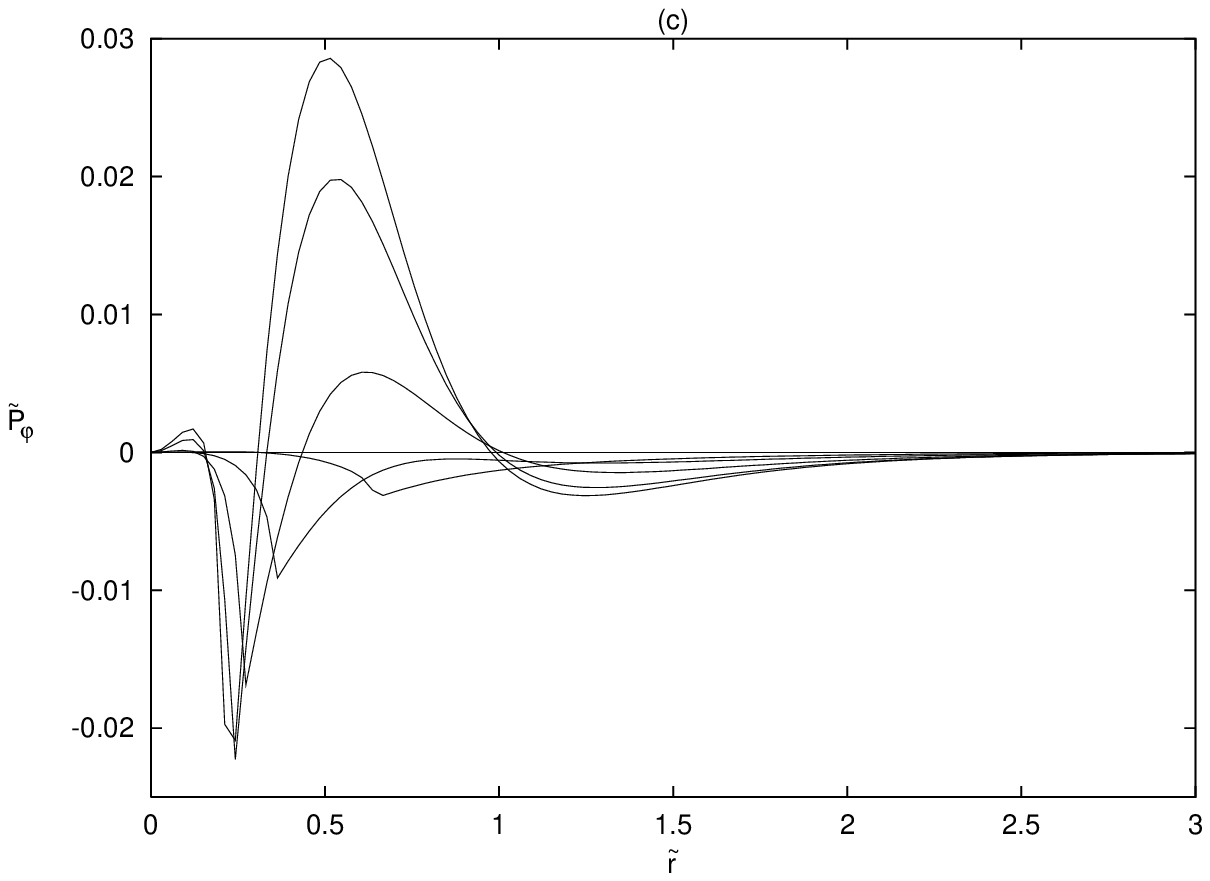}%
\includegraphics[scale=0.5]{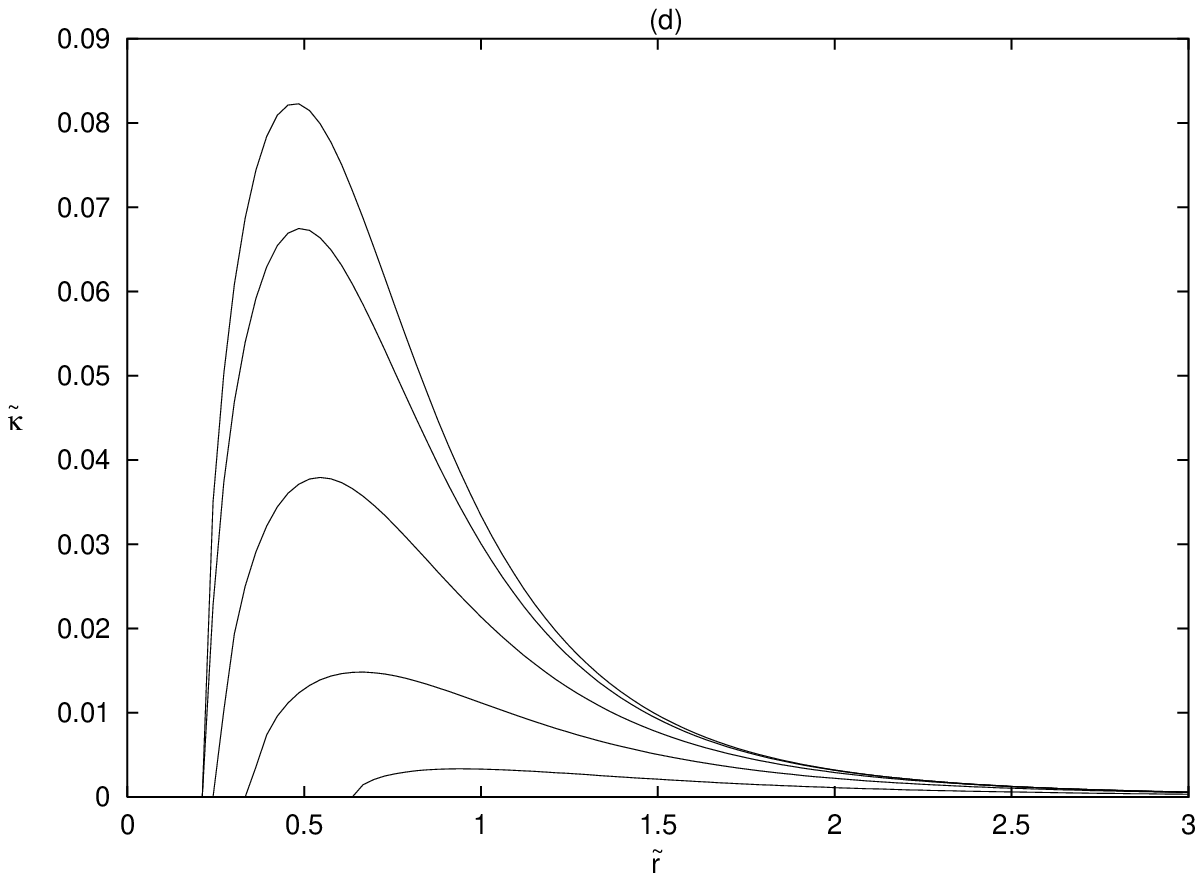}\\
\vspace{0.2cm}
\includegraphics[scale=0.6]{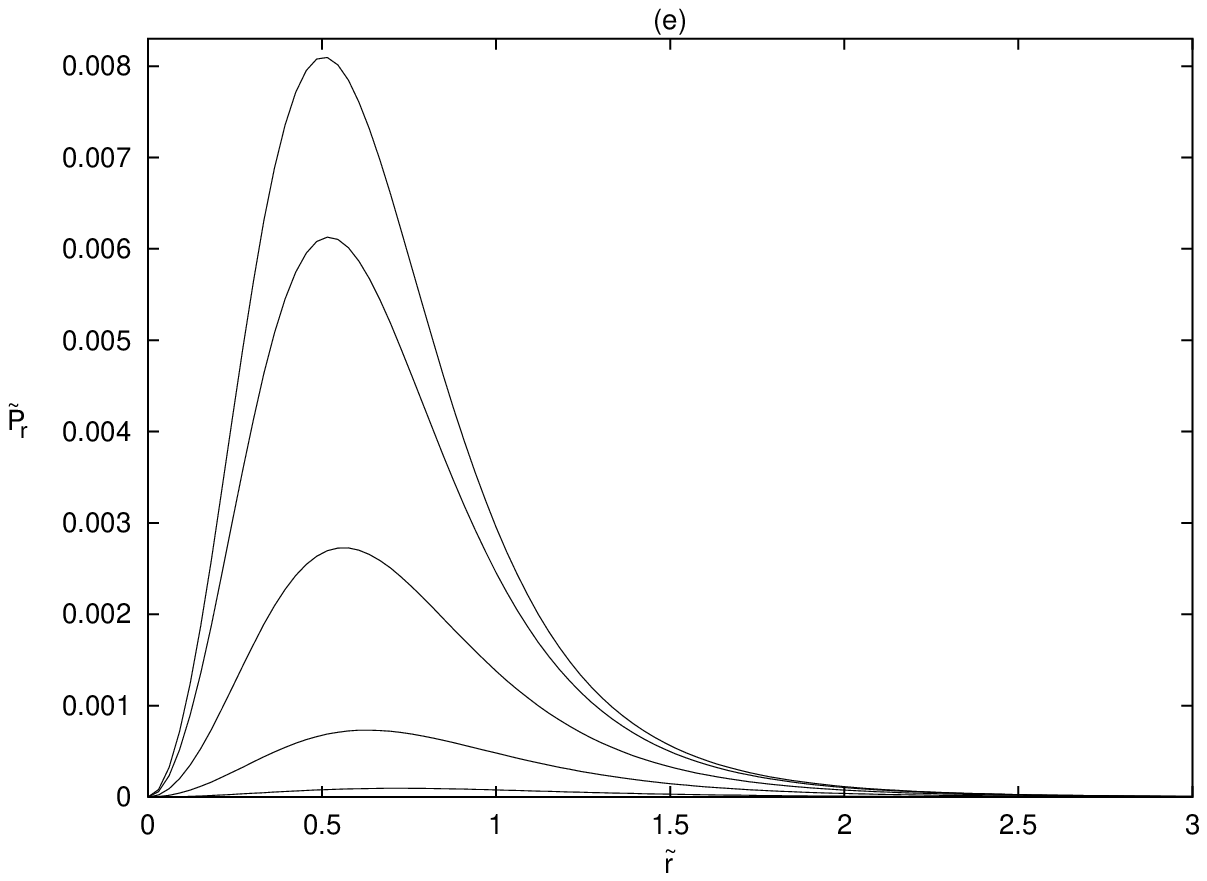}
\caption{(a) The discriminant Eq.\ (\ref{eq_D_bon}), (b) the energy density Eq.\ (\ref{eq_rho_disc}), 
(c) the azimuthal stress Eq.\ (\ref{eq_Pphi_disc}), (d) the heat flow function (\ref{eq_heat_disc}) and (e) the 
radial pressure Eq.\ (\ref{eq_Pr}) as functions of $\tilde{r}$ for $\tilde{z}=0,0.2, 0.4, 0.6, 0.8,1$ and 
parameters $n=1$, $\tilde{k}=0.5$, $\tilde{d}=0$ and $\tilde{b}=1$. In (a) the small plot shows the curves 
for $\tilde{z}=0.8$ and $\tilde{z}=1$.} \label{fig2}
\end{figure}

Fig.\ \ref{fig2}(a)--(e) show, respectively, curves of the discriminant Eq.\ (\ref{eq_D_bon}), 
the energy density Eq.\ (\ref{eq_rho_disc}), the azimuthal stress Eq.\ (\ref{eq_Pphi_disc}), 
the heat flow function (\ref{eq_heat_disc}) and the radial pressure Eq.\ (\ref{eq_Pr}). The parameters 
used are $n=1$, $\tilde{k}=0.5$, $\tilde{d}=0$ and $\tilde{b}=1$. Each curve is a cut taken on the planes $\tilde{z}=0$ (the 
curves with largest amplitudes), $\tilde{z}=0.2, 0.4, 0.6, 0.8,1$. From Fig.\ \ref{fig2}(b) we have an 
annular region of the disk with negative energy density. For the azimuthal stress (Fig.\ \ref{fig2}(c)) 
there are two rings with pressure (positive stress), however near the vertical borders of the disk we 
only have tension. Most of the disk radial extension has heat flow (Fig.\ \ref{fig2}(d)) that is zero on $\tilde{z}=1$. From 
Fig.\ \ref{fig2}(e) we see that the radial pressure (as well as tension along $\tilde{z}$) is concentrated in the central regions of the disk. 
\section{Thick disks from a ``monopole'' and ``quadrupole'' solution} \label{sec_quad}

\begin{figure}
\centering
\includegraphics[scale=0.8]{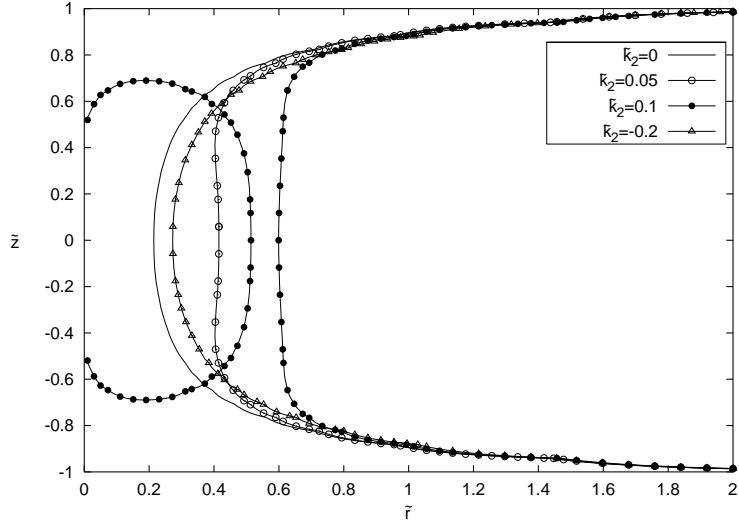}
\caption{Curves of $\tilde{D}=0$ as function of $\tilde{r}$ and $\tilde{z}$ for $\tilde{k}_2=0,0.05,0.1,-
0.2$. Parameters: $n=1$, $\tilde{d}=0$, $\tilde{k}_1=1$ and $\tilde{b}=1$.} \label{fig3}
\end{figure}

We consider another solution of Eq.\ (\ref{eq_Phi}), namely, a sum of a monopole-like and 
a quadrupole-like term 
\begin{equation} 
\Phi=\frac{k_1}{\sqrt{r^2+z^2}}+\frac{k_2(r^2-2z^2)}{(r^2+z^2)^{5/2}} \mbox{,}
\end{equation}
where $k_1$ and $k_2$ are constants (note that $k_1=2k$ when compared with Bonnor's solution). 
Eqs.\ (\ref{eq_stockum2})--(\ref{eq_stockum3}) and the 
relation $N=r\Phi_{,r}$ give
\begin{subequations}
\begin{align}
N &=-\frac{r^2}{(r^2+z^2)^{7/2}}\left[ k_1(r^2+z^2)^2+3k_2(r^2-4z^2) \right] \mbox{,} \label{eq_N_quad} \\
\nu &=\frac{r^2}{16(r^2+z^2)^8} \left[ 2k_1^2(r^2+z^2)^4 (r^2-8z^2) +24k_1k_2(r^2+z^2)^2 \times \right. \notag \\
& \left. (r^4-18r^2z^2+16z^4) +9k_2^2(9r^6-288r^4z^2+672r^2z^4-256z^6) \right] \mbox{,} \\
\rho &=\frac{c^2}{8\pi G e^{\nu}(r^2+z^2)^8} \left[k_1^2(r^2+z^2)^4 (r^2+4z^2)+6k_1k_2(r^2+z^2)^2 \times \right. \notag \\
& \left. (3r^4+12r^2z^2-16z^4) +9k_2^2(9r^6+72r^4z^2-48r^2z^4+64z^6) \right] \mbox{.}
\end{align}
\end{subequations}
Applying again the transformation $z \rightarrow h(z)+b$ on the above solutions, we obtain for the 
physical quantities of the disk exact but long expressions that are not very illuminating. In Fig.\ \ref{fig3} 
we show how the Eq.\ $\tilde{D}=0$ is changed with the parameter $\tilde{k}_2=k_2/a^4$ (all other quantities were adimensionalised as in Sec.\ \ref{sec_bon}), for $\tilde{h}$ with $n=1$, $\tilde{d}=0$, 
$\tilde{k}_1=1$ and $\tilde{b}=1$. Values for $\tilde{k}_2$ somewhat greater then $0.053$ give rise 
to a second level curve where $\tilde{D}=0$.   

\begin{figure}
\centering
\includegraphics[scale=0.5]{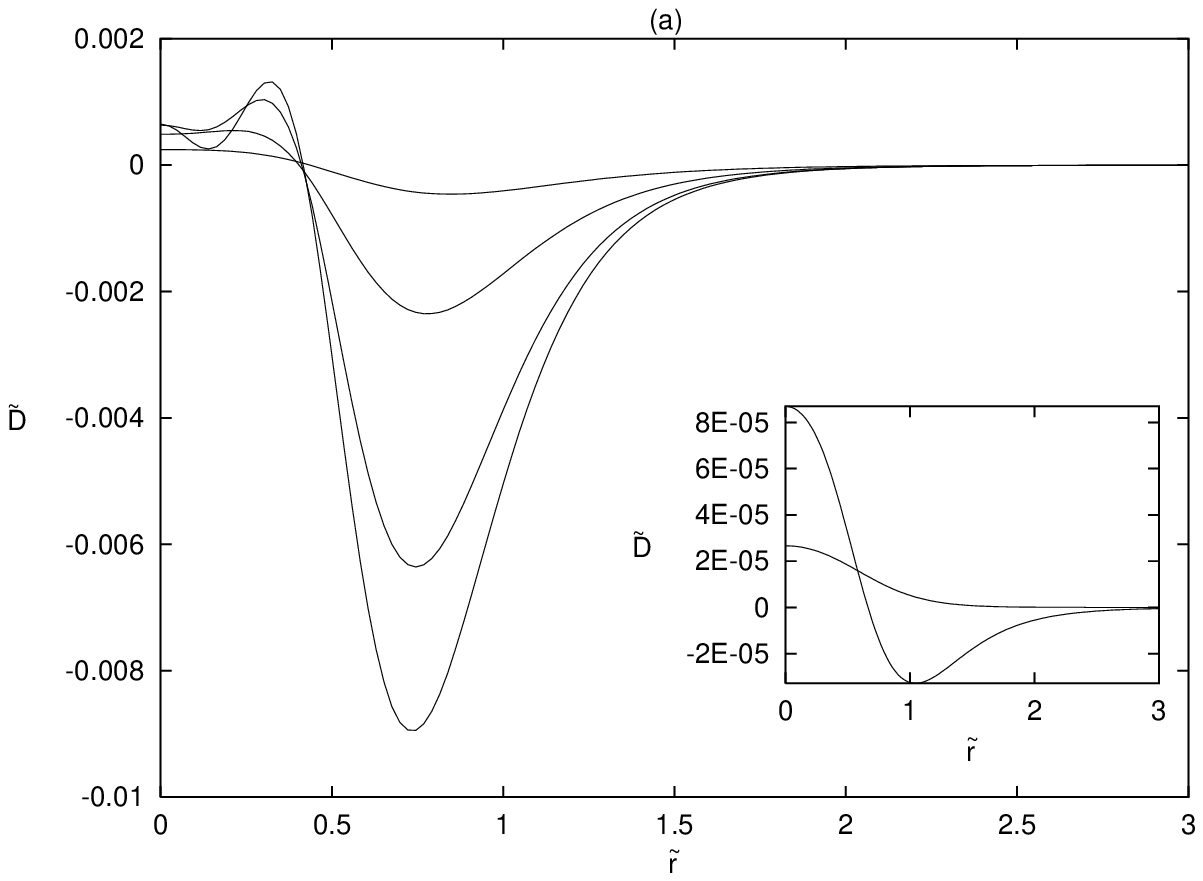}%
\includegraphics[scale=0.5]{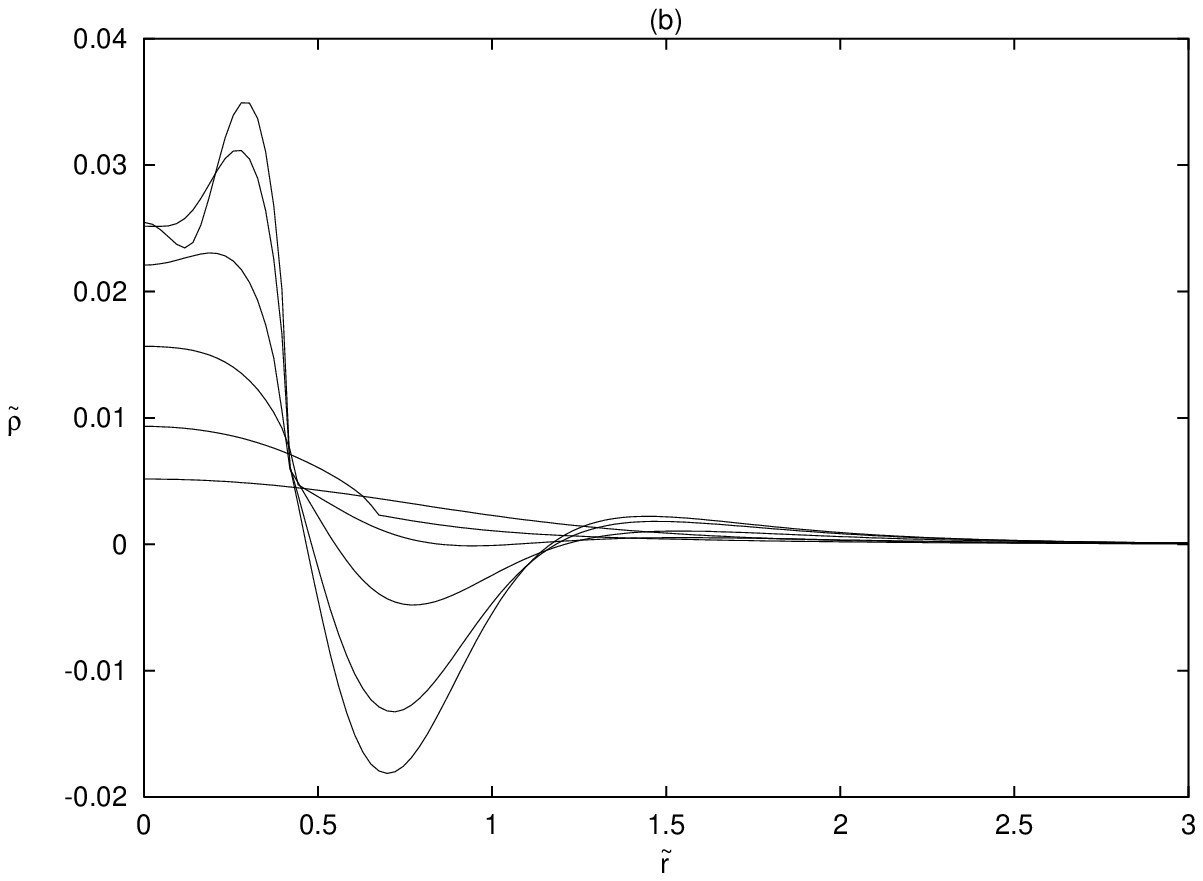}\\
\vspace{0.2cm}
\includegraphics[scale=0.5]{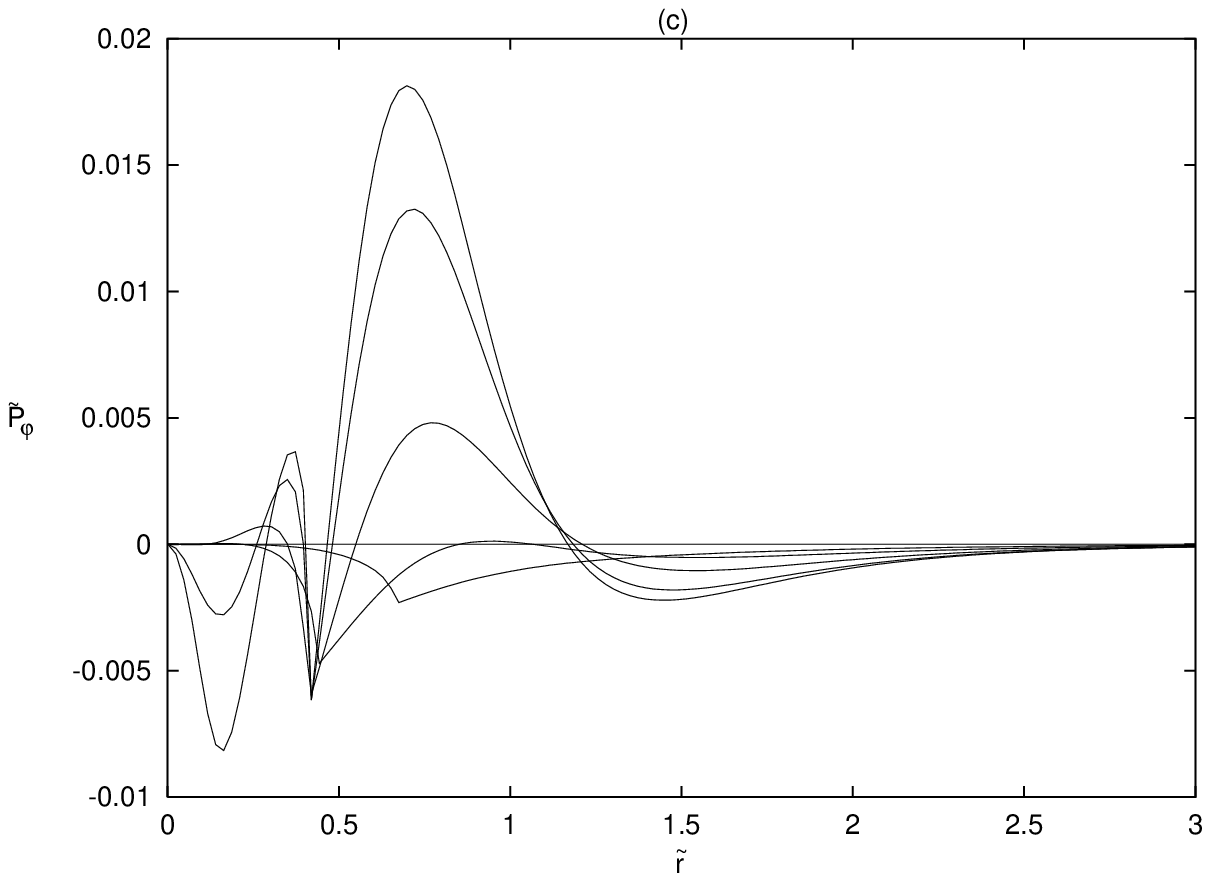}%
\includegraphics[scale=0.5]{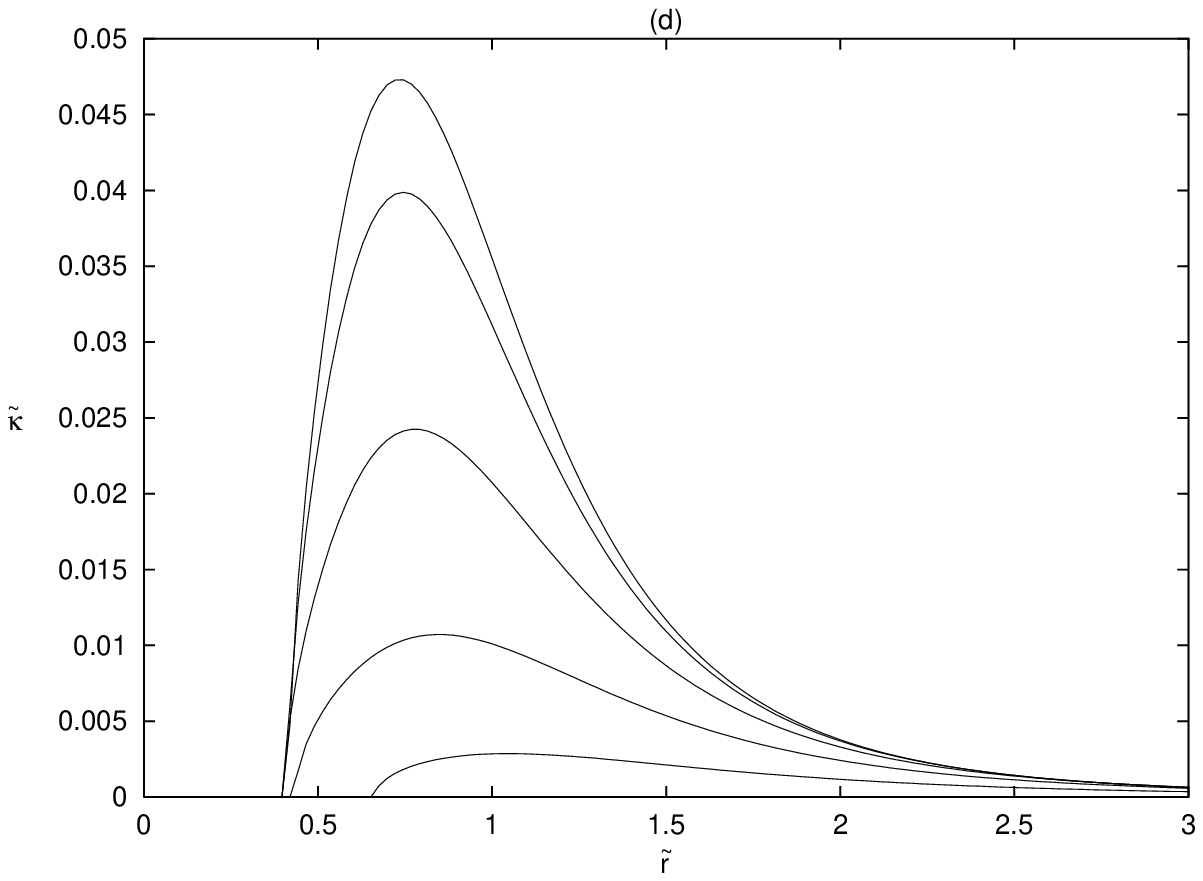}\\
\vspace{0.2cm}
\includegraphics[scale=0.5]{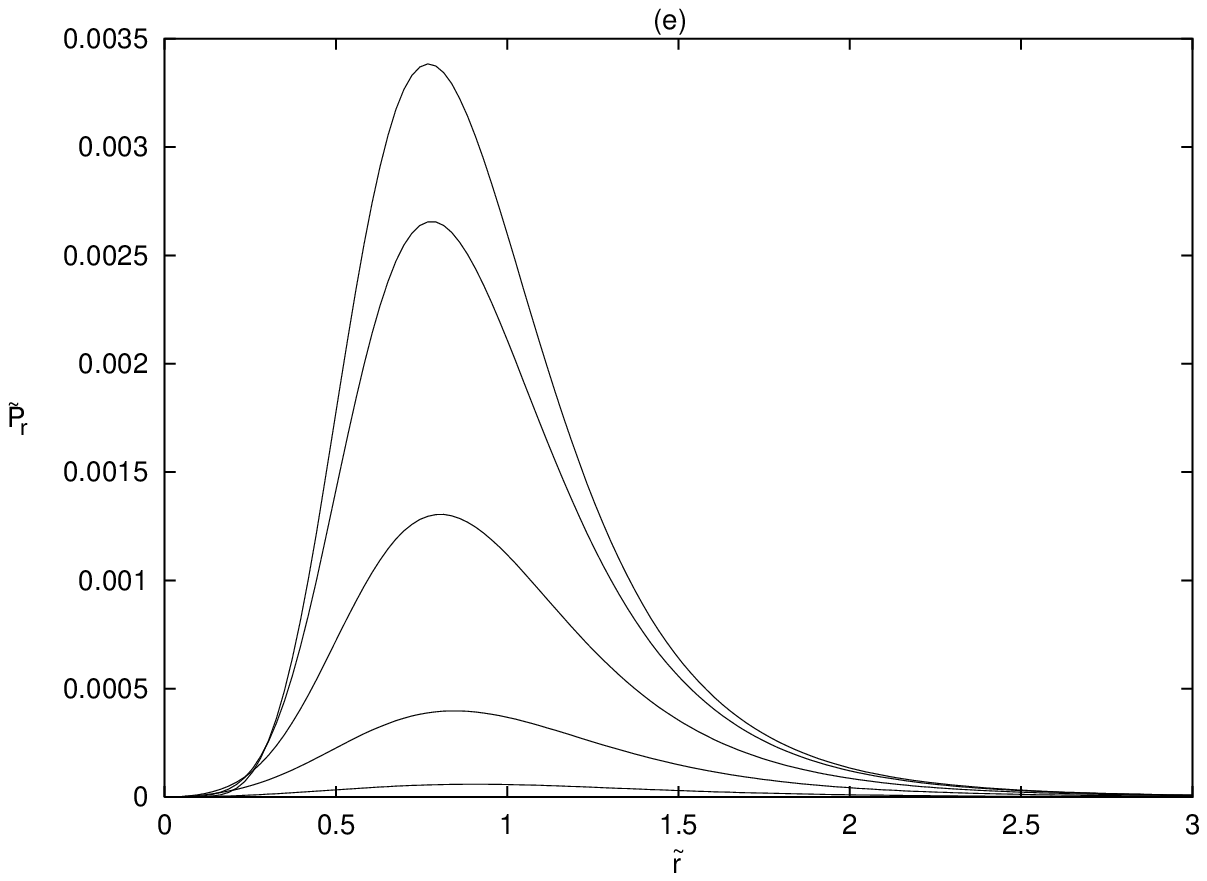}
\caption{(a) The discriminant $\tilde{D}$ (b) the energy density Eq.\ (\ref{eq_rho_disc}), 
(c) the azimuthal stress Eq.\ (\ref{eq_Pphi_disc}), (d) the heat flow function (\ref{eq_heat_disc}) and (e) the 
radial pressure Eq.\ (\ref{eq_Pr}) as functions of $\tilde{r}$ for $\tilde{z}=0,0.2, 0.4, 0.6, 0.8,1$ and 
parameters $n=1$, $\tilde{k}_1=1$, $\tilde{d}=0$ and $\tilde{b}=1$. In (a) the small plot shows the curves 
for $\tilde{z}=0.8$ and $\tilde{z}=1$.} \label{fig4}
\end{figure} 
Fig.\ \ref{fig4}(a)--(e) display, respectively, curves of the discriminant $\tilde{D}$
the energy density Eq.\ (\ref{eq_rho_disc}), the azimuthal stress Eq.\ (\ref{eq_Pphi_disc}), 
the heat flow function (\ref{eq_heat_disc}) and the radial pressure Eq.\ (\ref{eq_Pr}). The parameters 
used are $n=1$, $\tilde{k}_1=1$, $\tilde{k}_2=0.05$, $\tilde{d}=0$ and $\tilde{b}=1$. As in Fig.\  \ref{fig2} each curve is a cut taken on the planes $\tilde{z}=0$ (the curves with largest amplitudes), $\tilde{z}=0.2, 0.4, 0.6, 0.8,1$. When compared with Fig.\ \ref{fig2}(a)--(e), the curves are shifted 
to the right. We also note the appearance of a new local maximum point in the energy density 
(Fig.\ \ref{fig4}(b)) and a new local minimum in the azimuthal stress (Fig.\ \ref{fig4}(c)). The curves for the heat flow function and radial pressure are not changed qualitatively.   
\section{Thin disks and the Cooperstock and Tieu galactic model} \label{sec_exotic}

The thin disk limit is achieved by choosing the function $h(z)=|z|$ in the 
transformation $z \rightarrow h(z)+b$. Since $h^{\prime}=2\theta(z)-1$ and 
$h^{\prime\prime}=2\delta(z)$, where $\theta(z)$ and $\delta(z)$ are, respectively, the 
Heaviside function and the Dirac distribution, the components (\ref{eq_Ttt2})--(\ref{eq_Trr_Tzz}) of 
the energy-momentum tensor will have a distributional part with support on  the plane $z=0$: 
$T^a_b=Q^a_b\delta(z)$. The nonzero components of $Q^a_b$ are
\begin{subequations}
\begin{align}
Q^t_t &=\frac{c^4}{8\pi Ge^{\nu}}\left(\frac{NN_{,h}}{r^2}-\nu_{,h} \right), \qquad 
Q^t_{\varphi}=-\frac{c^4N_{,h}}{8\pi Ge^{\nu}}\left(1+\frac{N^2}{r^2} \right) \mbox{,} \label{eq_Qtt}\\
Q^{\varphi}_{t} &=\frac{c^4N_{,h}}{8\pi G r^2e^{\nu}}, \qquad 
Q^{\varphi}_{\varphi} =-\frac{c^4}{8\pi G e^{\nu}} \left(\frac{NN_{,h}}{r^2}+\nu_{,h} \right) \mbox{,} \label{eq_Qphiphi}
\end{align}
\end{subequations}
The calculation of the eigenvalues result in
\begin{align}
S &= Q^t_t+ Q^{\varphi}_{\varphi}=-\frac{c^4\nu_{,h}}{4\pi Ge^{\nu}} \mbox{,} \\
D &= (Q^t_t- Q^{\varphi}_{\varphi})^2+4 Q^t_{\varphi}Q^{\varphi}_t= -\frac{c^8
 N_{,h}^2}{16\pi^2G^2r^2e^{\nu}}<0 \mbox{.}
\end{align}
Because the discriminant is always negative, the surface energy density $\sigma$ of the thin disk is given by 
$\sigma=S/(2c^2)$ and the azimuthal stress by $P_{\varphi}=-S/2$. Thus the disk is composed of matter 
with an equation of state $P_{\varphi}=-c^2\sigma$. If $\sigma>0$ we have tensions and it may 
be interpreted as an equation of state of matter formed by concentric loops of cosmic strings \cite{Let79}. 
If $\sigma<0$ we have pressure and matter
with negative energy density. Objects with this equation of state are known in the literature as struts and they appear to stabilize 
certain superpositions of static isolated bodies in General Relativity (see, for instance, \cite{Bach22}). In 
either case the thin disks are made of exotic matter with unusual properties. Note that this result is 
general in the sense that it is independent of particular forms for the metric functions $N$ and $\nu$.

In the galactic model presented by Cooperstock and Tieu \cite{Coop05} the relevant field equations that describe the distribution 
of galactic matter are, to order $G$, given by
\begin{align}
& N_{,rr} +N_{,zz}-\frac{N_{,r}}{r}=0,  \label{eq_N}\\
\rho &=\frac{c^2}{8\pi Gr^2}\left( N_{,r}^2+N_{,z}^2 \right)  \mbox{.} \label{eq_rho_G}
\end{align}
 The metric function $N(r,z)$ is calculated from the measured tangential velocity $V=cN/r=c\Phi_{,r}$, and then the density distribution is determined by the nonlinear relation (\ref{eq_rho_G}). A convenient set of basis functions that are a solution of equation (\ref{eq_N}) is given by
\begin{equation} \label{eq_N_serie}
N=-\sum_n C_nk_ne^{-k_n|z|}rJ_1(k_nr) \mbox{,}
\end{equation}
where $J_1$ is the Bessel function of order 1 and the $C_n$ are constants. The absolute value of $z$ must 
be used to ensure the reflexive symmetry with respect to the plane $z=0$. However, this is equivalent to make the transformation $z \rightarrow |z|$ and by the results stated above, introduces on the plane $z=0$ an additional rotating disk made of exotic matter. It is worth to note that in another context
 de Ara\'ujo and Wang \cite{Wang00} have considered a solution of equation (\ref{eq_N}) in the form $e^{-|z|}rJ_1(r)$ and also comment that this introduces an additional mass layer with an equation of state $P_{\varphi}=-c^2\sigma$ and with heat flow. 

In principle the exact solution (\ref{eq_N_serie}) may also be used to generate thick disks via the 
transformation $z \rightarrow h(z)$ (with $b=0$), however the calculation of the other metric function $\nu$ 
using relations (\ref{eq_stockum2}) is far from trivial. But at least the zeros of the discriminant equation 
(\ref{eq_D}) can be determined straightforwardly, since they do not depend on the function $\nu$. 
\begin{figure}
\centering
\includegraphics[scale=0.8]{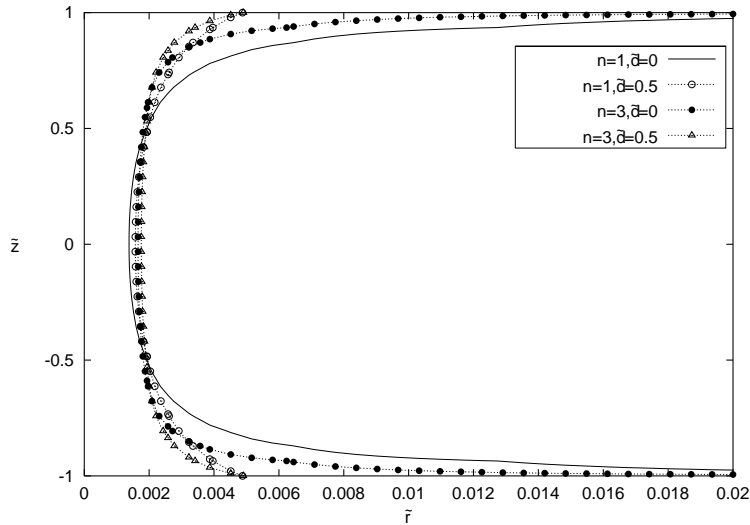}
\caption{Curves of $\tilde{D}=0$ as function of $\tilde{r}$ and $\tilde{z}$ for $h$ with $n=1$ and $n=3$. 
The curves were calculated with Eq.\ (\ref{eq_N_milky}) fitted for the Milky Way (table 1 of \cite{Coop05}).} \label{fig5}
\end{figure}
As an example we take $N$ as 
\begin{equation} \label{eq_N_milky}
N=-a\sum_{n=1}^{10}C_nk_ne^{-ak_n\tilde{h}(z)}\tilde{r}J_1(ak_n\tilde{r}) \mbox{,}
\end{equation}
where the $C_nk_n$ and $k_n$ were fitted for the Milky Way as given in Table 1 of  \cite{Coop05}. Fig.\ 
\ref{fig5} shows some curves of $\tilde{D}=0$ for $h$ with $n=1$ and $n=3$. The variables were adimensionalised as in Sec.\ \ref{sec_bon} and the half-thickness of the Milky Way was set to $a=0.75$ 
Kpc. On the left of the curves we have small intervals of radii where $\tilde{D}>0$. In the limit 
$a \rightarrow 0$ these intervals would shrink to zero, leaving a singular layer on $z=0$ with exotic matter.   
\section{Discussion} \label{sec_discuss}

Using the van Stockum class of metrics that describe spacetimes with stationary axially symmetric 
distributions of dust, and applying the ``displace, cut, fill and reflect'' method, we constructed thick rotating disks surrounded by rotating dust. In general the disks have equal radial pressures and tensions in the $z$ 
direction. The solution found by Bonnor \cite{Bon77} and a sum of a ``monopole'' and ``quadrupole'' solution were 
used to study particular models of rotating disks. They were found to have annular regions with negative energy 
density and azimuthal pressure. Heat flow is concentrated near the plane $z=0$ and is present in most 
of the radial extension of the disk. These particular examples suggest that the 
``displace, cut, fill and reflect'' method used on the van Stockum class of metrics will generate 
disks with some unrealistic physical properties and thus are hardly useful in galactic modelling. However, the aplication of the above mentioned method 
 to solutions obtained using a different
approach (see, for instance, \cite{Bal06}) may result in more realistic and useful models.

In the thin limit we obtain disks with heat flow everywhere and composed of exotic matter with an equation 
of state $P_{\varphi}=-c^2\sigma$, which may be interpreted as an equation of state of cosmic strings or 
struts. We also comment the new galactic model proposed by Cooperstock and Tieu \cite{Coop05} and the presence of an 
additional singular layer of exotic matter in their model.

\section*{Acknowledgments}
D.\ Vogt thanks CAPES for financial support. P.\ S.\ Letelier thanks CNPq and FAPESP for financial support.

\end{document}